\documentstyle[psfig]{mn2e}

\def\erg{{\rm\thinspace erg}}
\def\ergps{{\rm\thinspace erg~s^{-1}}}
\def\yr{{\rm\thinspace yr}}

\def\etal{{\it et al.\ }}
\def\eg{{\it e.g.\ }}

\def\spose#1{\hbox to 0pt{#1\hss}}
\def\approxlt{\mathrel{\spose{\lower 3pt\hbox{$\sim$}}
	\raise 2.0pt\hbox{$<$}}}
\def\approxgt{\mathrel{\spose{\lower 3pt\hbox{$\sim$}}
	\raise 2.0pt\hbox{$>$}}}
\def\approxpropto{\mathrel{\spose{\lower 3pt\hbox{$\sim$}}
	\raise 2.0pt\hbox{$\propto$}}}
\mathchardef\twiddle="2218

\def\multleft#1{\hbox to size{\vbox {\halign {\lft{##}\cr #1}}\hfill}\par}
\def\multright#1{\hbox to size{\vbox {\halign {\rt{##}\cr #1}}\hfill}\par}

\def\today{\ifcase\month\or January\or February\or March\or April\or May\or
      June\or July\or August\or September\or October\or November\or December\fi
      \space\number\day, \number\year}
\def\<{\thinspace}

\def\erg{{\rm\thinspace erg}}

\def\km{{\rm\thinspace km}}

\def\Mpc{{\rm\thinspace Mpc}}
\def\Msun{\hbox{$\rm\thinspace M_{\odot}$}}

\def\s{{\rm\thinspace s}}
\def\yr{{\rm\thinspace yr}}

\def\ergps{\hbox{$\erg\s^{-1}\,$}}

\def\kmps{\hbox{$\km\s^{-1}\,$}}

\def\Msunpyr{\hbox{$\Msun\yr^{-1}\,$}}

\def\kmpspMpc{\hbox{$\kmps\Mpc^{-1}$}}

\title[The relation between accretion rate and jet power 
in elliptical galaxies]{The relation between accretion rate 
and jet power in X-ray luminous elliptical galaxies}

\author[S.W. Allen \etal ]
{\parbox[]{6.in} {S.W. Allen$^1$, R.J.H. Dunn$^2$, A.C. Fabian$^2$,
G.B. Taylor$^3$ and C.S. Reynolds$^4$ \\
\footnotesize
1. Kavli Institute for Particle Astrophysics and Cosmology, Stanford University, 382 Via Pueblo Mall, Stanford, CA 94305-4060, USA.  \\
2. Institute of Astronomy, Madingley Road, Cambridge CB3 0HA. \\
3. University of New Mexico, Department of Physics and Astronomy, Albuquerque, NM 87131, USA. \\
4. Department of Astronomy, University of Maryland, College Park, MD 20742, USA.}}
\begin{document}
\maketitle
\begin{abstract}
Using Chandra X-ray observations of nine nearby, X-ray luminous
elliptical galaxies with good optical velocity dispersion
measurements, we show that a tight correlation exists between the
Bondi accretion rates calculated from the observed gas temperature and
density profiles and estimated black hole masses, and the power
emerging from these systems in relativistic jets. The jet powers,
which are inferred from the energies and timescales required to
inflate cavities observed in the surrounding X-ray emitting gas, can
be related to the accretion rates using a power law model of the form
log\,($P_{\rm Bondi}/10^{43}\ergps) = A + B\,${\rm log}\,($P_{\rm
jet}/10^{43}\ergps)$, with $A=0.65\pm0.16$ and $B=0.77\pm0.20$. Our
results show that a significant fraction of the energy associated with
the rest mass of material entering the Bondi accretion radius
($2.2^{+1.0}_{-0.7}$ per cent, for $P_{\rm jet}=10^{43}$\ergps)
eventually emerges in the relativistic jets. The data also hint that this
fraction may rise slightly with increasing jet power.  Our results
have significant implications for studies of accretion, jet formation
and galaxy formation.  The observed tight correlation suggests that
the Bondi formulae provide a reasonable description of the accretion
process in these systems, despite the likely presence of magnetic
pressure and angular momentum in the accreting gas. The similarity of
the $P_{\rm Bondi}$ and $P_{\rm jet}$ values argues that a significant
fraction of the matter entering the accretion radius flows down to
regions close to the black holes, where the jets are presumably
formed. The tight correlation between $P_{\rm Bondi}$ and $P_{\rm
jet}$ also suggests that the accretion flows are approximately stable
over timescales of a few million years. Our results show that the
black hole `engines' at the hearts of large elliptical galaxies and
groups can feed back sufficient energy to stem cooling and star formation,
leading naturally to the observed exponential cut off at the bright
end of the galaxy luminosity function.
\end{abstract}

\begin{keywords} accretion, accretion disks -- black hole physics -- 
galaxies:active -- galaxies:jets -- X-rays: galaxies
\end{keywords}

\section{Introduction}

The black holes in the hearts of galaxies and galaxy clusters are
commonly observed to be associated with powerful relativistic
jets. The mechanism by which such jets form and the efficiency with
which the energy associated with material entering the accretion
radius is converted into jet power at much smaller radii remains the
subject of much debate.  Such knowledge is important for understanding
the nature of the accretion process, galaxy formation and the growth
of supermassive black holes.

X-ray studies with Chandra and XMM-Newton, building on earlier work
with ROSAT, have shown that the black holes at the centres of
galaxies, groups and clusters interact strongly with their
environments, blowing clear cavities or `bubbles' in the surrounding
X-ray emitting gas (\eg B\"ohringer \etal 1993; Fabian \etal 2000,
2003, 2005; Birzan \etal 2004; Forman \etal 2005). Simple arguments
based on the energy required to inflate the bubbles and the sound
speed of the surrounding gas allow these bubbles to be used as
calorimeters, measuring the power released by the jets (\eg Churazov
\etal 2002). Importantly, the exquisite spatial resolution of Chandra
means that for the largest, nearby ellipticals we are also able to
resolve regions close to the Bondi accretion radii in these systems.
Given estimates for the central black hole masses from measured galaxy
velocity dispersions, the observed density and temperature profiles of
the X-ray emitting gas can be used to estimate the accretion rates
onto the black holes (Bondi 1952).  Taken together, the accretion
rates and jet powers then allow us to estimate the efficiency with
which the rest mass of accreted matter is converted into jet power.

The observed luminosities of AGN in elliptical galaxies are typically
several orders of magnitude lower than would be predicted from the
standard Bondi formulae, assuming a conversion efficiency for matter
into radiation of order 10 per cent. However, much of the predicted
power may actually be released in the form of jets. This issue was
investigated by Di Matteo \etal (2003) and Taylor \etal (2006) who
noted that although the central AGN in NGC4486 (M87), NGC4696 and
NGC6166 have bolometric luminosities three orders of magnitude below
the Bondi predictions (for 10 per cent efficiency), the jet powers
inferred from the observed X-ray cavities are comparable to the Bondi
values, within a factor of a few. Pellegrini \etal (2003) used related
arguments to infer a minimum conversion efficiency for accreted mass
into jet power of $\sim 1$ per cent for the elliptical galaxy IC4296.
Here, we extend the work of Di Matteo \etal 2003 and Taylor \etal
(2006) to a sample of nine nearby, X-ray bright ellipticals with good
X-ray data and optical velocity dispersion measurements.

The origin of the observed galaxy luminosity function and, in
particular, the mechanism responsible for truncating star formation in
the largest ellipticals is also a subject of much debate (\eg Benson
\etal 2003, Croton \etal 2005; Bower \etal 2005, and references
therein). Here, again, significant insight has been provided by
XMM-Newton and Chandra observations (\eg Tamura \etal 2001; Allen
\etal 2001; Peterson \etal 2001, 2003) which show that the cooling
rates in the cores of galaxies, groups and clusters are much smaller
than the values predicted by simple models that do not include
feedback. Some widespread form of heating must be operating in
galaxies and clusters to stem cooling. We show here that the 
kinetic power released in jets from the central black hole engines can, in
principle, provide this heat and stem further star formation.

\section{observations and data analysis}

\subsection{Target selection}

Our target galaxies are bright, nearby ellipticals with accurate
central black hole mass measurements or velocity dispersions which
allow the black hole masses to be estimated \eg using the relation of
Tremaine \etal (2002).  The Chandra archive was searched to identify
all such systems with X-ray luminous gaseous halos and obvious
jet-induced cavities in the X-ray emitting gas. We have restricted our
target list to those systems in which the bubbles are attached (or
very close to) to the central AGN, which are therefore likely to be
undergoing inflation.  We have also required that the Chandra exposure
times be sufficient to allow precise measurements of the central
density and temperature of the X-ray emitting gas. Finally, we
required that the X-ray data allow us to measure the properties of the
X-ray emitting gas within one order of magnitude of the Bondi radius,
permitting reliable measurements or extrapolations of the gas
properties at that radius to be made. These selection criteria are
matched by eight galaxies with data in the Chandra archive.  For
comparison purposes, we also include results for one more distant
system, NGC6166, the dominant galaxy of Abell 2199. Although, due to
its greater distance, we do not come close to resolving the Bondi
radius in this system, the density profile follows a simple power-law
form which allows an interesting extrapolation to smaller radii.

\subsection{Chandra analysis}

\begin{table}
\begin{center}
\caption{Summary of the Chandra observations. Columns list the target
name (and alternative), distance in Mpc, observation date and net
exposure after all cleaning and screening procedures were carried out.
For the Virgo Cluster ellipticals, a distance of 17 Mpc has been
assumed.  For NGC 507, 4696, 5846 and 6166 the distances are luminosity
distances for a flat $\Lambda$CDM cosmology with $\Omega_{\rm m}=0.3$
and a Hubble Constant $H_0=70$\kmpspMpc.}\label{table:obs}
\begin{tabular}{ c c c c c  }
                & $d_L$ (Mpc)  &  Date    & Exposure (ks) \\
\hline	            
NGC507          & 71.4   &   2000 Oct 11  & 19.1  \\
NGC4374 (M84)   & 17     &   2000 May 19  & 28.1  \\
NGC4472         & 17     &   2001 Jun 12  & 21.9   \\
NGC4486 (M87)   & 17     &   2000 Jul 29  & 33.7  \\
NGC4552 (M89)   & 17     &   2001 Apr 22  & 53.4  \\
NGC4636         & 17     &   2000 Jan 26  & 43.2  \\
NGC4696         & 44.9   &   2004 Apr 01  & 86.7  \\ 
NGC5846         & 24.6   &   2000 May 24  & 19.5  \\
NGC6166         & 135.5   &   1999 Dec 11  & 15.0  \\
\hline                      
\end{tabular}
\end{center}
\end{table}

\begin{table*}
\begin{center}
\caption{Summary of the measured Bondi accretion rates and powers. 
Columns list the galaxy velocity dispersion ($\sigma$ in units
km\,s$^{-1}$), the logarithm of the accretion radius
($r_{\rm A}$ in units of parsecs), 
slope of the inner electron density profile
($\alpha$), electron density at the accretion radius ($n_{\rm
e}$($r=r_{\rm A}$) in units cm$^{-3}$), temperature at the accretion radius 
($kT$($r=r_{\rm A}$) in units of keV), logarithm of the Bondi accretion rate 
(${\dot M}_{\rm Bondi}$ in units of \Msunpyr), and the logarithm of the 
predicted accretion power ($P_{\rm Bondi}$ in units of $10^{43}$erg\,s$^{-1}$). 
Uncertainties are 68 per cent confidence
limits. Additional systematic uncertainties
of 0.23 dex in log\,$r_{\rm A}$, and 0.46 dex in 
log\,${\dot M}_{\rm Bondi}$ and log\,$P_{\rm Bondi}$, originating from 
the observed systematic scatter in the 
log\,$M_{\rm BH}-$log\,$\sigma$ relation (Tremaine \etal 2002), 
may be associated with these values. See text for details.}
\label{table:bondi}
\vskip 0 truein
\begin{tabular}{ c c c c c c c c r  }
               & ~ & log\,$\sigma$    &  log\,$r_{\rm A}$      & $\alpha$            & $n_{\rm e}$($r=r_{\rm A}$) & $kT$($r=r_{\rm A}$) & log\,${\dot M}_{\rm Bondi}$& log\,$P_{\rm Bondi}$\\
\hline	       	     									                            
NGC507         & ~ &  $2.498\pm0.014$ & $1.58_{-0.09}^{+0.09}$ & $-1.10^{+0.04}_{-0.04}$ & $3.23^{+1.03}_{-0.71}$ & $0.74\pm0.04$ & $-1.34_{-0.09}^{+0.09}$  & $1.41_{-0.09}^{+0.09}$    \\
NGC4374 (M84)  & ~ &  $2.473\pm0.010$ & $1.49_{-0.07}^{+0.08}$ & $-0.55^{+0.19}_{-0.18}$ & $0.94^{+0.84}_{-0.46}$ & $0.71\pm0.05$ & $-2.07_{-0.29}^{+0.30}$  & $0.69_{-0.29}^{+0.30}$    \\
NGC4472        & ~ &  $2.488\pm0.013$ & $1.56_{-0.12}^{+0.14}$ & $-0.36^{+0.12}_{-0.12}$ & $0.83^{+0.42}_{-0.28}$ & $0.70\pm0.16$ & $-1.96_{-0.23}^{+0.25}$  & $0.79_{-0.23}^{+0.25}$    \\
NGC4486 (M87)  & ~ &  $2.557\pm0.013$ & $2.09_{-0.20}^{+0.13}$ & $-0.00^{+0.10}_{-0.10}$ & $0.17^{+0.04}_{-0.03}$ & $0.80\pm0.01$ & $-1.59_{-0.40}^{+0.28}$  & $1.16_{-0.40}^{+0.28}$    \\
NGC4552 (M89)  & ~ &  $2.410\pm0.031$ & $1.26_{-0.14}^{+0.14}$ & $-0.83^{+0.08}_{-0.08}$ & $1.30^{+0.76}_{-0.45}$ & $0.67\pm0.09$ & $-2.38_{-0.21}^{+0.22}$  & $0.37_{-0.21}^{+0.22}$    \\
NGC4636        & ~ &  $2.318\pm0.017$ & $0.99_{-0.11}^{+0.12}$ & $-0.31^{+0.09}_{-0.09}$ & $0.42^{+0.22}_{-0.15}$ & $0.54\pm0.11$ & $-3.46_{-0.24}^{+0.24}$  &$-0.71_{-0.24}^{+0.24}$   \\
NGC4696        & ~ &  $2.418\pm0.014$ & $1.22_{-0.07}^{+0.07}$ & $-0.63^{+0.30}_{-0.30}$ & $1.61^{+3.82}_{-1.15}$ & $0.81\pm0.05$ & $-2.35_{-0.55}^{+0.56}$  & $0.40_{-0.55}^{+0.56}$    \\ 
NGC5846        & ~ &  $2.416\pm0.012$ & $1.29_{-0.08}^{+0.09}$ & $-0.35^{+0.26}_{-0.21}$ & $0.35^{+0.52}_{-0.22}$ & $0.67\pm0.09$ & $-2.90_{-0.40}^{+0.43}$  &$-0.15_{-0.40}^{+0.43}$   \\
NGC6166        & ~ &  $2.496\pm0.012$ & $1.31_{-0.20}^{+0.31}$ & $-0.50^{+0.07}_{-0.06}$ & $0.92^{+0.41}_{-0.33}$ & $1.3\pm0.7$   & $-2.27_{-0.26}^{+0.34}$  & $0.49_{-0.26}^{+0.34}$    \\
\hline                      
\end{tabular}
\end{center}
\end{table*}

The Chandra observations were carried out using the Advanced CCD
Imaging Spectrometer (ACIS) between 1999 December and 2004 April. The
ACIS-S array was used as the primary detector as it offers the best
sensitivity to soft X-ray emission from the galaxies. The standard
level-1 event lists produced by the Chandra pipeline processing were
reprocessed using the $CIAO$ (version 3.2.1) software package,
including the latest gain maps and calibration products. Bad pixels
were removed and standard grade selections applied.  Where possible,
extra information available in VFAINT mode was used to improve the
rejection of cosmic ray events. The data were cleaned to remove
periods of anomalously high background using the recommended energy
ranges and time binning. The net exposure times after cleaning are
summarized in Table~\ref{table:obs}. The analyses of NGC4486 (M87) and
NGC6166 have been reported separately by Di Matteo \etal (2001, 2003).
For these objects, we have used their density and temperature
profiles, scaled to the distances summarized in Table~\ref{table:obs}.

The temperature and density profiles of the X-ray emitting gas were
determined using the methods described by Allen \etal (2004 and
references therein). In brief, concentric annular spectra were
extracted from the cleaned event lists, centred on the AGN (identified
from radio data and/or the presence of an obvious X-ray point source
in the Chandra data). These spectra were analysed using XSPEC
(version 11.3: Arnaud 1996), the MEKAL plasma emission code (Kaastra
\& Mewe 1993; incorporating the Fe-L calculations of Liedhal,
Osterheld \& Goldstein 1995) and the photoelectric absorption models
of Balucinska-Church \& McCammon (1992). 
We have included standard
correction factors to account for time-dependent contamination along
the instrument light path and have incorporated a small correction to
the High Resolution Mirror Assembly model in CIAO 3.2.1, which takes
the form of an 'inverse' edge with an energy, E=2.08\,keV and optical
depth $\tau=-0.1$ (Herman Marshall, private communication).  Only data
in the $0.6-8.0$ keV energy range were used for our analysis. The
spectra for all annuli for a particular galaxy were modelled
simultaneously in order to determine the deprojected X-ray gas
temperature profiles, under the assumption of spherical symmetry.
\footnote{Note that for NGC4696, 
we have only analysed data from a 35 degree wide slice, covering 
position angles 15-50 degrees. This excludes the complex structure to 
the west of the central source discussed by Taylor \etal (2006). 
For all other sources, the full 360 degrees of position angle
were used. Note that for NGC4486, the bright 
non-thermal emission from the jets was also excluded 
(Di Matteo \etal 2003)}.  The
emission from each spherical shell was modelled as a single phase
plasma, with the abundances of all metals in each shell 
assumed to vary with a
common ratio, $Z$, with respect to Solar values. (The
exception to the latter assumption was the analysis of NGC4486,
reported separately by Di Matteo \etal (2003), for which the
abundances of key elements were permitted to vary independently in
each shell.)

Background spectra were extracted from the blank-field data sets
available from the Chandra X-ray Center. These were cleaned in an
identical manner to the target observations.  In each case, the
normalizations of the background files were scaled to match the count
rates in the target observations measured in the 9.5-12keV band.
Separate photon-weighted response matrices and effective area files
were constructed for each region using calibration files appropriate
for the period of observation.

Azimuthally-averaged surface brightness profiles were constructed from
background subtracted, flat-fielded images with a $0.492\times0.492$
arcsec$^2$ pixel scale ($1\times1$ raw detector pixels). Together with
the deprojected spectral temperature profiles, these were used to
determine the X-ray gas density profiles (see \eg White, Jones \& Forman
1997 for details of the technique).

\section{Calculation of accretion rates and jet powers}

\subsection{Black hole masses and accretion radii}

\noindent For the case of M87, the measurement of the central black
hole mass, $M_{\rm BH}=3.0\pm1.0\times10^9$\Msun, is from Tremaine
\etal (2002) and is based on the data of Harms \etal (1994) and
Macchetto \etal (1997). For the other galaxies in the sample, the
black hole masses have been estimated using the correlation between
$M_{\rm BH}$ and velocity dispersion, $\sigma$, given by
Tremaine \etal (2002)

\begin{equation}
{\rm log}\,(M_{\rm BH}/M_{\sun}) = \alpha + \beta( {\rm log}\, \sigma/200\kmpspMpc), 
\end{equation}

\noindent with $\alpha=8.13\pm0.06$ and $\beta=4.02\pm0.32$. Our
analysis uses a Monte Carlo algorithm to account for the 
uncertainties in $\sigma$ and the  
slope of the log\,$M_{\rm BH}-$log\,$\sigma$ relation. 
Velocity dispersions for the
galaxies are drawn from the work of Bernardi \etal (2002) and are
summarized in Table~\ref{table:bondi}.

The accretion radius, $r_{\rm A}$, is the radius within which the
gravitational potential of the central black hole dominates over the
thermal energy of the surrounding X-ray emitting gas

\begin{equation}
r_{\rm A} = 2 G M_{\rm BH}/c_s^2.
\end{equation}

\noindent Here $G$ is the gravitational constant, $c_s=\sqrt{\gamma_1
kT/\mu m_p}$ is the adiabatic sound speed of the gas at the accretion
radius, $T$ is the gas temperature at that radius, $\mu=0.62$ is the
mean atomic weight, $m_p$ is the proton mass and $\gamma_1$ is the
adiabatic index of the X-ray emitting gas.  The results on the
accretion radii, summarized in Table~\ref{table:bondi}, account for
the uncertainties in the central black hole masses (described above) 
and gas temperatures.  

Tremaine \etal (2002) measure an intrinsic (systematic) dispersion of
0.23 dex in the log\,$M_{\rm BH}-$log\,$\sigma$ relation, which
implies a similar systematic uncertainty in the log\,$r_{\rm A}$
values.  This systematic uncertainty in $r_{\rm A}$ dominates over the
tabulated measurement errors in most cases. The effects of accounting
for this systematic dispersion are discussed in
Section~\ref{section:relation}.

\begin{figure*}
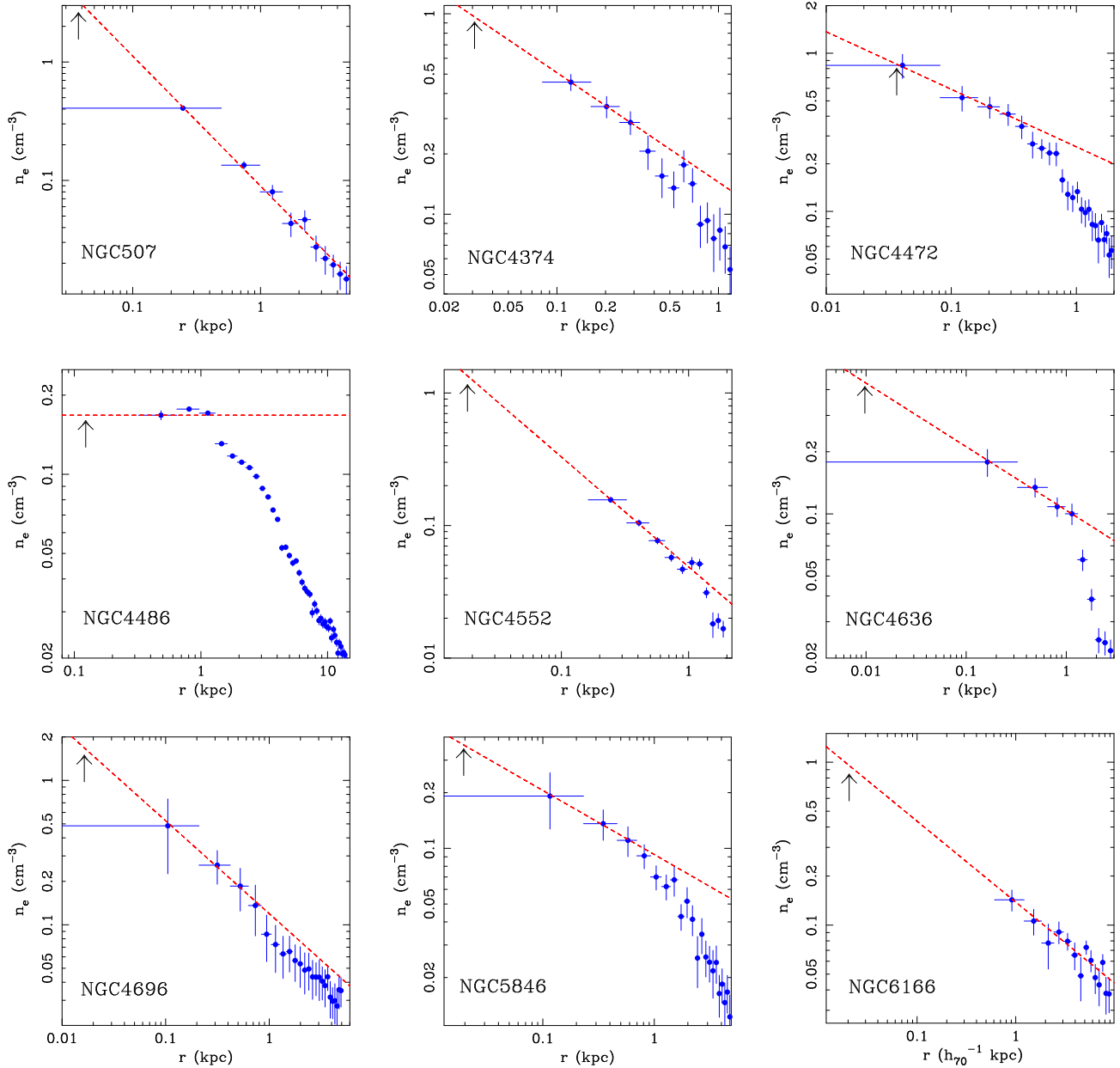

\vspace{0.5cm}
\hbox{
\hspace{0.0cm}\psfig{figure=ne_ngc507.ps,width=.30\textwidth,angle=270}
\hspace{0.5cm}\psfig{figure=ne_ngc4374.ps,width=.30\textwidth,angle=270}
\hspace{0.5cm}\psfig{figure=ne_ngc4472.ps,width=.30\textwidth,angle=270}
} 
\vspace{0.5cm}
\hbox{
\hspace{0.0cm}\psfig{figure=ne_m87.ps,width=.30\textwidth,angle=270}
\hspace{0.5cm}\psfig{figure=ne_ngc4552.ps,width=.30\textwidth,angle=270}
\hspace{0.5cm}\psfig{figure=ne_ngc4636.ps,width=.30\textwidth,angle=270}
} 
\vspace{0.5cm}
\hbox{
\hspace{0.0cm}\psfig{figure=ne_ngc4696.ps,width=.30\textwidth,angle=270}
\hspace{0.5cm}\psfig{figure=ne_ngc5846.ps,width=.30\textwidth,angle=270}
\hspace{0.5cm}\psfig{figure=ne_ngc6166.ps,width=.30\textwidth,angle=270}
} 
\vspace{0.5cm}
\caption{The observed electron density profiles for the central
regions of the galaxies, as determined from the Chandra data. For NGC
4374, 4486, 4552 and 6166 the data from the innermost radii, which are
contaminated by non-thermal emission from the central AGN, have been
excluded.  The dashed lines show the best-fitting power law models
($n_e(r)=n_1r^{\alpha}$) which have been fitted to the inner few bins;
the number of bins fitted are noted in parentheses, after the galaxies
names below. The Bondi accretion radius, $r_{\rm A}$, for each galaxy
is marked with an arrow.  From top to bottom, left to right, we show
NGC507(10), 4374(3), 4472(4), 4486(1), 4552(3), 4636(4), 4696(4),
5846(3) and 6166(14).}\label{fig:ne}
\end{figure*}

\begin{figure*}
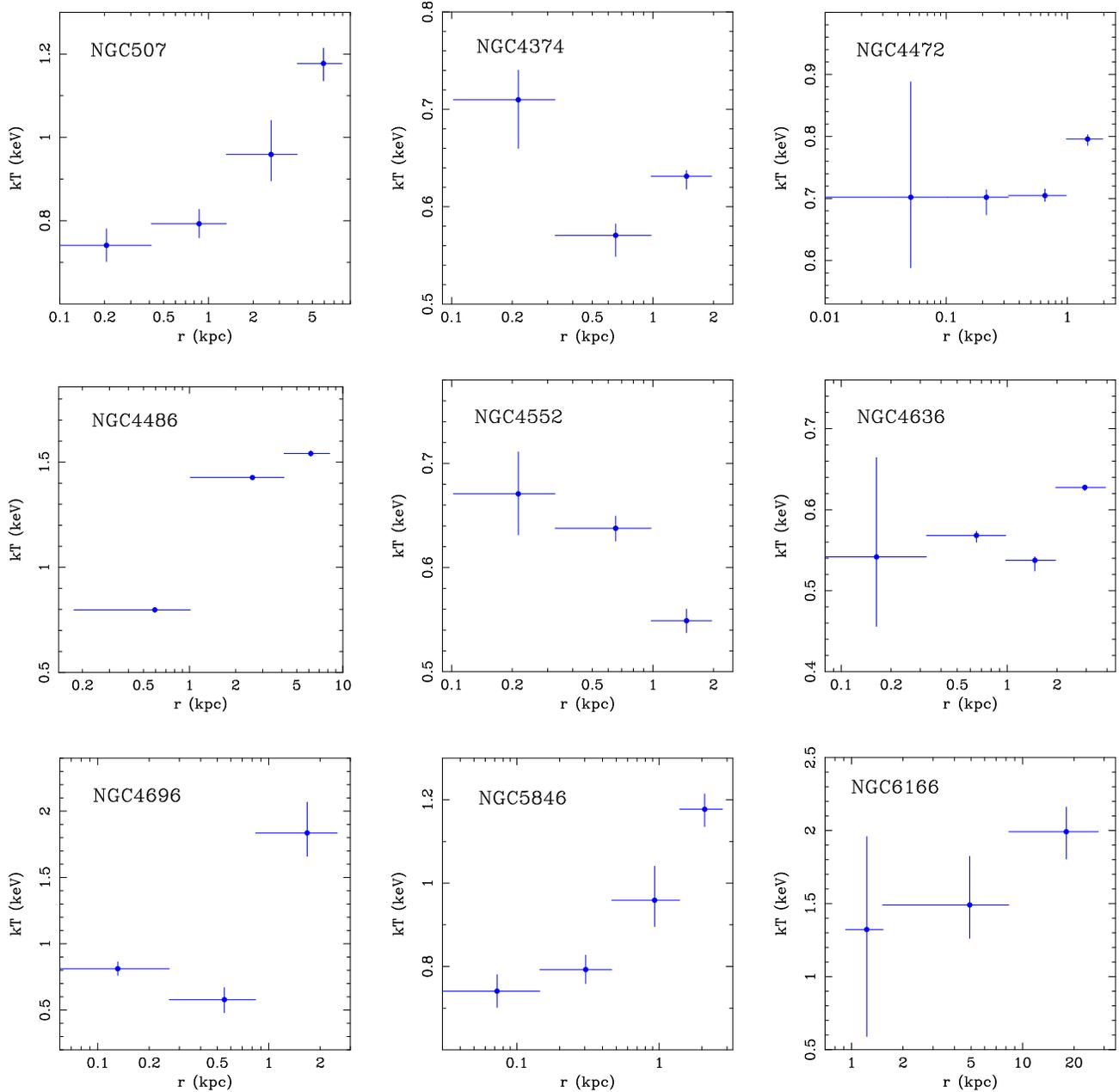

\vspace{0.5cm}
\hbox{
\hspace{0.0cm}\psfig{figure=kt_ngc507.ps,width=.30\textwidth,angle=270}
\hspace{0.5cm}\psfig{figure=kt_ngc4374.ps,width=.30\textwidth,angle=270}
\hspace{0.5cm}\psfig{figure=kt_ngc4472.ps,width=.30\textwidth,angle=270}
} 
\vspace{0.5cm}
\hbox{
\hspace{0.0cm}\psfig{figure=kt_m87.ps,width=.30\textwidth,angle=270}
\hspace{0.5cm}\psfig{figure=kt_ngc4552.ps,width=.30\textwidth,angle=270}
\hspace{0.5cm}\psfig{figure=kt_ngc4636.ps,width=.30\textwidth,angle=270}
} 
\vspace{0.5cm}
\hbox{
\hspace{0.0cm}\psfig{figure=kt_ngc4696.ps,width=.30\textwidth,angle=270}
\hspace{0.5cm}\psfig{figure=kt_ngc5846.ps,width=.30\textwidth,angle=270}
\hspace{0.5cm}\psfig{figure=kt_ngc6166.ps,width=.30\textwidth,angle=270}
} 
\vspace{0.5cm}
\caption{The X-ray gas temperature profiles for the central regions of the
galaxies determined from the Chandra data.}\label{fig:kt}
\end{figure*}


\begin{figure*}
\vspace{0.5cm}
\hbox{
\hspace{0.0cm}\psfig{figure=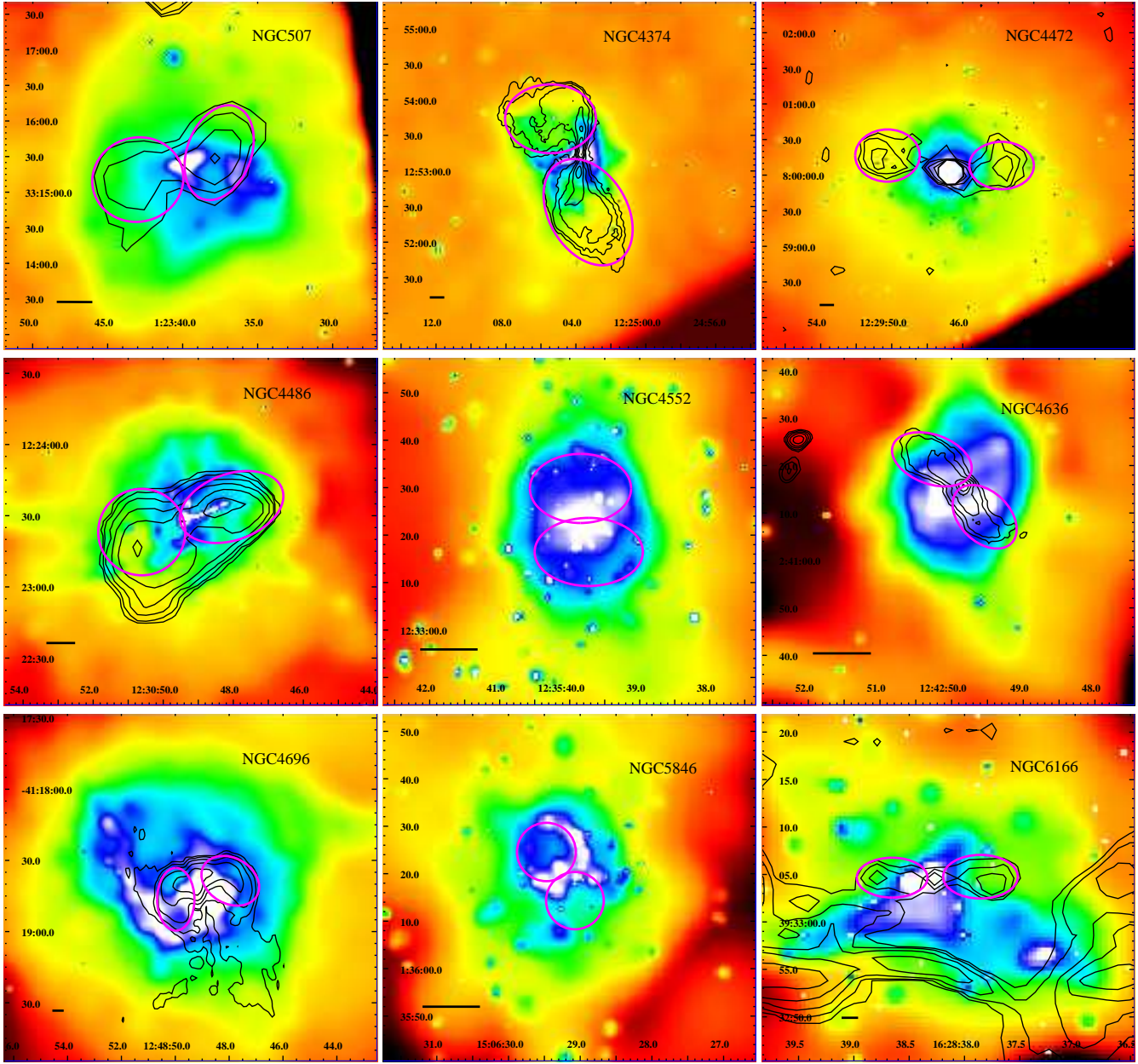,width=1.0\textwidth,angle=0}
} 
\vspace{0.5cm}
\caption{Chandra X-ray images for the central regions of the galaxies.
The X-ray images are constructed in the 0.5-8 keV band and have been
adaptively smoothed using 2 or 3-pixel Gaussian kernels. The regions
identified with the X-ray cavities, which are used to estimate the jet
powers, are indicated by bold magenta contours.  The narrow black
contours show the 1.5 GHz radio emission from archival Very Large
Array (VLA) data.  The radio data help to identify the cavities in
cases where identifications from the X-ray data alone may be
ambiguous.  The radio contours are logarithmically spaced and have
been adjusted to highlight the extended lobe emission. The horizontal
scale bar in the bottom left corner of each image denotes a length of
1kpc, except for NGC507 where the scale is
10kpc.}\label{fig:img}
\end{figure*}



\begin{table*}
\begin{center}
\caption{Summary of the measured jet powers. In order, we list the distances $R$ from the black holes to the 
centres of the X-ray bubbles, the energy $E$ required to expand the bubbles, the bubble ages 
$t_{\rm age}$, and the power required to inflate the bubbles over this timescale $P_{\rm jet}$.
The quoted uncertainties in $R$ include statistical measurement errors only. A factor 2 
systematic uncertainty in the bubble volumes has been included in the calculation of 
$E$ and $P_{\rm jet}$.}\label{table:buoy}
\vskip 0.1 truein
\begin{tabular}{ l c c c c c }
\hline
  Galaxy    &Lobe  &  $R$            &$E=4pV$            &$t_{\rm age}$  & $P_{\rm jet}$     \\
            &      &  (kpc)          &($10^{54}\erg$)    &($10^6 \yr$)    & ($10^{43} \ergps$) \\
\hline	       	       							                       
NGC507      & E    & $12.8\pm1.0 $   &  $48900\pm21100$  & $23.6\pm2.13 $ &  $  6.60\pm2.91 $   \\
            & W    & $11.3\pm1.0 $   &  $29300\pm13300$  & $25.6\pm2.18 $ &  $  3.62\pm1.70 $   \\
NGC4374     & N    & $3.03\pm0.3 $   &  $ 1650\pm  693$  & $ 6.07\pm0.56$ &  $ 0.868\pm0.371$   \\
            & S    & $4.36\pm0.4 $   &  $ 2050\pm  854$  & $10.0\pm1.08 $ &  $ 0.658\pm0.268$   \\
NGC4472     & E    & $4.33\pm0.4 $   &  $  730\pm  287$  & $ 5.28\pm0.56$ &  $ 0.445\pm0.181$   \\
            & W    & $3.40\pm0.3 $   &  $  606\pm  244$  & $ 5.29\pm0.57$ &  $ 0.362\pm0.150$   \\
M87         & W    & $1.86\pm0.05$   &  $ 1210\pm  782$  & $ 2.92\pm0.13$ &  $  1.26\pm0.84$    \\
            & E    & $1.51\pm0.05$   &  $ 1670\pm 1100$  & $ 2.38\pm0.11$ &  $  2.18\pm1.44 $   \\
NGC4552     & N    & $0.58\pm0.02$   &  $ 35.0\pm 12.7$  & $ 1.53\pm0.08$ &  $ 0.0727\pm0.0274$ \\
            & S    & $0.58\pm0.02$   &  $ 39.4\pm 14.3$  & $ 1.51\pm0.08$ &  $ 0.0829\pm0.0307$ \\
NGC4636     & NE   & $0.69\pm0.07$   &  $ 8.10\pm 3.21$  & $ 1.77\pm0.15$ &  $ 0.0147\pm0.0060$ \\
            & SW   & $0.66\pm0.07$   &  $ 8.04\pm 3.25$  & $ 1.68\pm0.17$ &  $ 0.0152\pm0.0060$ \\
NGC4696     & E    & $2.49\pm0.3 $   &  $  601\pm  338$  & $ 6.06\pm0.66$ &  $ 0.313\pm0.177$   \\
            & W    & $3.00\pm0.3 $   &  $  826\pm  438$  & $ 5.54\pm0.60$ &  $ 0.478\pm0.244$   \\
NGC5846     & N    & $0.72\pm0.12$   &  $ 21.0\pm 11.2$  & $ 1.81\pm0.30$ &  $ 0.0367\pm0.0185$ \\
            & S    & $0.72\pm0.12$   &  $ 20.8\pm 11.3$  & $ 1.81\pm0.30$ &  $ 0.0374\pm0.0188$ \\
NGC6166     & E    & $2.30\pm0.10$   &  $ 829 \pm 336 $  & $ 3.51\pm0.27$ &  $ 0.741\pm0.316$   \\
            & W    & $2.40\pm0.10$   &  $ 981 \pm 383 $  & $ 3.66\pm0.28$ &  $ 0.838\pm0.353$   \\
\hline                      
&&&&&\\
\end{tabular}
\end{center}
\end{table*}

\subsection{X-ray gas properties at the accretion radii}

Fig~\ref{fig:ne} shows the electron density profiles for the central
regions of the galaxies, determined from the Chandra X-ray data. In
cases where the X-ray emission from the innermost parts is
contaminated by non-thermal emission from a central AGN (NGC4374,
4486, 4552, 6166), the affected regions have been excluded.  The
arrows in the figures mark the locations of the accretion radii. With
the exception of NGC\,4472, the accretion radii are unresolved by the
Chandra data. However, in general, the inner few points of the density
profiles can be described by a power-law model of the form
$n_e(r)\propto r^{\alpha}$ with $-1.1<\alpha <0$
(Table~\ref{table:bondi}).  We have used the power-law models to
estimate the gas densities at the accretion radii, in each case
accounting for uncertainties in the slopes and normalizations.  For
M87 (NGC4486), the density profile flattens within $r\sim 10r_{\rm A}$
and so we have assumed that the density at $r_{\rm A}$ is equal to the
value at the innermost measurement radius. (We assign a systematic
uncertainty in the slope $\alpha=0.0\pm0.1$ in this case.) Note that
the regions fitted with the power law density models are the 
regions for which we expect the assumption of spherical symmetry to be 
most valid, based on the Chandra images. 

The temperature profiles for the inner regions of the galaxies are
shown in Fig.~\ref{fig:kt}. In general, the profiles do not exhibit
steep gradients within $r\sim 10-100r_{\rm A}$. We have therefore
assumed that the temperatures at the accretion radii can be estimated
from the values at the innermost measurement radii. The temperature
results are summarized in Table~\ref{table:bondi}. Systematic
uncertainties in the temperature estimates are unlikely to affect
other results reported here significantly.

\subsection{Calculation of the accretion rates}

The calculation of accretion rates onto the central black holes
follows the work of Bondi (1952). Similar calculations are described
by \eg Di Matteo \etal (2001, 2003), Churazov \etal (2002), 
Pellegrini \etal (2003) and Taylor \etal (2006).

Under the assumption of spherical symmetry and negligible angular
momentum, the rate of accretion of the X-ray emitting gas at the
accretion radius can be written as (Bondi 1952)

\begin{equation}
{\dot M_{\rm Bondi}} = 4 \pi \lambda (GM_{\rm BH})^2 c_s^{-3}\rho
                     = \pi \lambda c_s \rho r_{\rm A}^2,
\end{equation}

\noindent where $\rho$ is the density of the gas at the accretion
radius (we assume $\rho=1.13 n_e m_p$) and $\lambda$ is a numerical
coefficient that depends upon the adiabatic index of the accreting
gas. For an assumed adiabatic index $\gamma_1=5/3$, $\lambda = 0.25$
(Bondi 1952). Note that for $\gamma_1=5/3$, $\rho/c_{\rm s}^3$ is
constant within the accretion flow, making it appropriate to use the
gas temperature and density measured at the accretion radius in our
calculations, in place of the values at infinity.  (In essence, we are
modelling the accretion flows as Bondi flows inward of the accretion
radii. For $r> r_{\rm A}$, the gravitational potentials of the
galaxies start to dominate and the Bondi formulae become
inappropriate.)

For an efficiency $\eta$, relating the accretion rate at the accretion
radius to the total energy emitted from within that radius, the
maximum power released from the black hole system is

\begin{equation}
P_{\rm Bondi} = \eta {\dot M_{\rm Bondi}} c^2. 
\end{equation}

\noindent The values of $P_{\rm Bondi}$, for an assumed value of
$\eta=0.1$ are summarized in Table ~\ref{table:bondi}. The 
statistical uncertainties in the results have been estimated from Monte Carlo
simulations, which account for the uncertainties in the $\sigma$
measurements, the slope of the log\,$M_{\rm BH}-$log\,$\sigma$ 
relation and uncertainties in the temperature and density measurements
at the accretion radii. 

An intrinsic (systematic) dispersion of 0.23 dex in the 
log\,$M_{\rm BH}-$log\,$\sigma$ relation (Tremaine \etal 2002) 
implies a systematic uncertainty of 0.46 dex in both 
log\,${\dot M_{\rm Bondi}}$ and log\,$P_{\rm Bondi}$.
These systematic errors are comparable to the 
measurement errors for most galaxies, and dominate for NGC507.

\subsection{Calculation of jet power}

To estimate the kinetic power of the jets, we first estimated the
energy, $E$ required to create the observed bubbles in the X-ray
emitting gas. For slow expansion rates, this is the sum of the
internal energy within the bubble and the $P{\rm d}V$ work done

\begin{equation}
E=\frac{1}{\gamma_2-1}PV+PV = \frac{\gamma_2}{\gamma_2-1}PV,
\end{equation}

\noindent where $P$ is the thermal pressure of the surrounding, X-ray
emitting gas (which can be determined from the observed X-ray
temperatures and densities), V is the volume of the cavity and
$\gamma_2$ is the mean adiabatic index of the fluid within the
cavity. For the case of a bubble filled with relativistic plasma,
$\gamma_2=4/3$ and $E=4PV$. Some additional energy will also be
transfered into sound waves (Churazov \etal 2002; Fabian \etal 2003,
2005, 2006; Forman \etal 2003; Ruszkowski \etal 2004).  We have neglected
this extra energy in our calculations, although the power involved is
likely to be smaller than the quoted uncertainties on the jet powers.

The regions identified with the bubbles are shown in
Fig.~\ref{fig:img}.  In general, the bubbles appear to be `attached'
to the central AGN and are likely to be undergoing expansion at
present. A possible exception to this is NGC4472, where the bubbles
may have recently detached. In cases where the edges of the bubbles
are not clearly defined in the X-ray data, we have used radio
observations to estimate their volumes (Fig.~\ref{fig:img}).  The
X-ray and radio data show that in general, the bubbles or cavities are
approximately elliptical. We have therefore parameterized the
projected areas of the bubbles in terms of a semi-axis length, $r_{\rm
l}$, along the jet direction and semi-axis width, $r_{\rm w}$, across
it. This allows the cavities to be modelled as ellipsoids with volumes
$V=4 \pi r_{\rm l} r_{\rm w} r_{\rm d}/3$, where $r_{\rm d}$ is the
unknown depth of the cavity along the line of sight.  We model the
cavities as prolate ellipsoids with $V=4\pi r_{\rm l} r_{\rm w}^2/3$,
although systematic uncertainties in $r_{\rm l}$, $r_{\rm w}$ and,
especially, $r_{\rm d}$ mean that our estimates of the bubble volumes
should be regarded as uncertain at the factor $\sim 2$ level.

Following Birzan \etal (2004; see also Dunn \& Fabian 2004, Dunn,
Fabian \& Taylor 2005) we have estimated the ages of the bubbles as

\begin{equation}
t_{\rm age}= R/c_s,
\end{equation}\label{eq:age}

\noindent where R is the distance of the bubble centre from the black
hole, which provides a reasonable match to some numerical simulations. 
We note that the bubbles have lower mass density than the surrounding
X-ray emitting gas and can therefore be expected to rise with a
buoyancy velocity, $v_{\rm b}=\sqrt{2gV/SC_D}$.  Here $S=\pi r_{\rm
w}^2$ is the cross-sectional area of the bubble (in the rise
direction), $V$ is the volume, $C_D\sim0.75$ is the drag coefficient
and $g=GM(<R))/R^2$ is the gravitational acceleration.  The times
required for bubbles of the observed sizes to rise buoyantly through a
uniform medium from the galaxy centres to their current radii is
approximately $t_{\rm buoy}=R/v_{\rm b}$. Application of this formula
leads to comparable timescales to the $t_{\rm age}$ values. However,
since the bubbles studied here are typically still attached to the
jets, we adopt $t_{\rm age}$ as our relevant timescale.

The power involved in `blowing' the bubbles can be estimated as

\begin{equation}
P_{\rm jet}=E/t_{\rm age}. 
\end{equation}

\noindent The results on $P_{\rm jet}$ are summarized in
Table~\ref{table:buoy}. The uncertainties on $P_{\rm jet}$ are
determined from Monte Carlo simulations which include the sources of
statistical uncertainty mentioned above as well as a factor 2
systematic uncertainty in the bubble volumes.

\begin{figure}
\vspace{0.5cm}
\hbox{
\hspace{0.0cm}\psfig{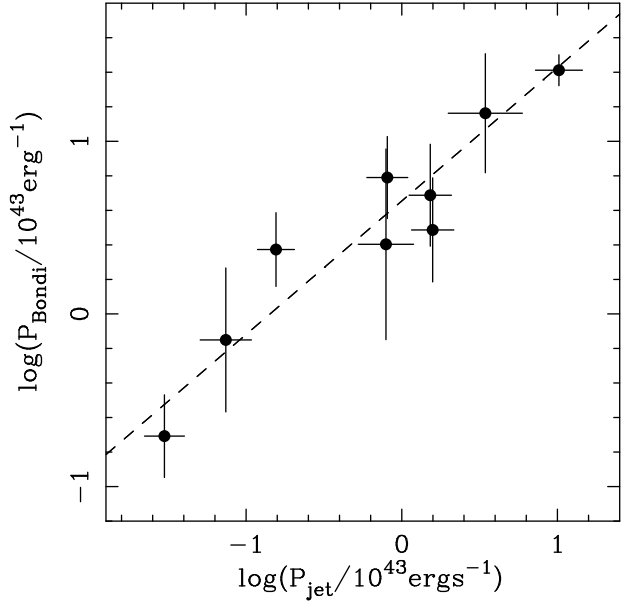}
} 
\caption{The logarithm of the Bondi accretion power 
(in units of $10^{43}$\ergps) 
determined from the Chandra 
X-ray data, assuming an efficiency
$\eta=0.1$ for the conversion of rest mass into energy, versus the logarithm of the 
jet power (also in units of $10^{43}$\ergps). The dashed line shows
the best-fitting linear-plus-constant model, determined using the BCES 
estimator. (See text for details.) 
}\label{fig:pbondi}
\end{figure}

\section{The relation between accretion rate and jet power}
\label{section:relation}

Fig~\ref{fig:pbondi} shows the results on the accretion power $P_{\rm
Bondi}$, for an assumed efficiency $\eta=0.1$ for the conversion of
accreted rest mass into energy, versus the jet power, $P_{\rm jet}$,
determined from the bubble properties. Both powers are in units of
$10^{43}$\ergps.  For each galaxy, we have combined the power
estimates for both bubbles, with the uncertainties added in
quadrature.  A clear correlation between the $P_{\rm Bondi}$ and
$P_{\rm jet}$ exists, which can be described by a power law model of
the form

\begin{equation}
{\rm log}\,\frac{P_{\rm Bondi}}{(10^{43}\ergps)}
= A + B {\rm log}\,\frac{P_{\rm jet}}{(10^{43}\ergps)}.  
\end{equation}

\noindent Using the $\chi^2$ fit statistic, which accounts only for
errors in the $P_{\rm Bondi}$ values, we find $A=0.67\pm0.07$ and
$B=0.74\pm0.08$,  with $\chi^2=5.3$
for 7 degrees of freedom (the quoted uncertainties on 
$A$ and $B$ are 68 per cent confidence intervals). 
Using the BCES($Y|X$) estimator of Akritas
\& Bershady (1996), which accounts for errors in both axes and the
presence of possible intrinsic scatter, we obtain $A=0.64 \pm 0.08$
and $B=0.78 \pm0.15$. The mean deviation about the best fitting BCES
model is $\sigma({\rm log}\,P_{\rm Bondi})=0.11$.

The results on $A$ and $B$ quoted above do not account for 
the effects of intrinsic (systematic) scatter in 
the log\,$M_{\rm BH}-$log\,$\sigma$ relation. We have 
used further Monte Carlo simulations to examine these
effects, introducing an intrinsic dispersion of 0.46 dex in
log\,$P_{\rm Bondi}$. Fits to the simulated data
sets using the BCES($Y|X$) estimator give $A=0.65\pm0.16$ and
$B=0.77\pm0.20$, in good agreement with the results presented above
but with (slightly) larger error bars on the fit parameters.

The results shown in Fig~\ref{fig:pbondi} indicate the presence of a strong
correlation between $P_{\rm Bondi}$ and $P_{\rm jet}$. A power-law 
model provides a good description of the data. However, in gauging the
origin and significance of this correlation, we must also 
consider effects that could arise 
from the plotted quantities having factors in common. 
Both $P_{\rm Bondi}$ and $P_{\rm jet}$ depend on the
distances to the objects. However, $P_{\rm Bondi}\propto d_{\rm L}^{-0.5}$, 
whereas $P_{\rm jet}\propto d_{\rm L}^{1.5}$. Thus, the distance
dependences cannot lead to the observed positive correlation. 
Both axes in Fig~\ref{fig:pbondi} also depend on the temperature $T$ of the
X-ray emitting gas.  However, the temperature shows little variation
from object to object and varies only mildly with radius between the
accretion and bubble radii. Moreover, the dependences on temperature
are approximately $P_{\rm Bondi}\propto T^{-1.5}$ and $P_{\rm
jet}\propto T^{1.5}$, which cannot lead to the observed 
positive correlation.  Finally, both axes involve the gas density,
measured at the accretion radius for $P_{\rm Bondi}$ and at the bubble
centres for $P_{\rm jet}$. These radii are very different and the
density profiles vary significantly from object to object, meaning
that the densities are essentially uncorrelated. (In detail, a mild
anti-correlation is observed). 

We conclude that the observed correlation between
$P_{\rm Bondi}$ and $P_{\rm jet}$ indicates a tight physical
connection between the two quantities.

\section{Implications and discussion}

We have shown that for supermassive black holes at the centres of
large, X-ray luminous elliptical galaxies, a remarkable, tight
correlation exists between the `Bondi' accretion rates inferred from
the Chandra X-ray data and observed galaxy velocity dispersions, and
the power emerging from these systems in relativistic jets.  Our
result has important implications for the nature of the accretion
process and for issues relating to feedback, the growth of black holes
and galaxy formation.
 
The relationship between the Bondi accretion power and jet power can
be described by a power law model of the form log\,$P_{\rm Bondi} =
0.65(\pm0.16) + 0.77(\pm0.20)${\rm log}$\,P_{\rm jet}$, where $P_{\rm
Bondi} = 0.1 {\dot M_{\rm Bondi}} c^2$ and $P_{\rm jet}$ is the power
associated with inflating the cavities and providing the internal
energy of the plasma that fills them. 
A significant fraction
($2.2^{+1.0}_{-0.7}$ per cent, for $P_{\rm jet}=10^{43}$\ergps) of the
energy associated with the rest mass of material entering the Bondi
accretion radius emerges from the systems in relativistic jets.  There
is a slight indication that this fraction increases as $P_{\rm jet}$
rises, from $1.3^{+1.0}_{-0.6}$ per cent at $P_{\rm
jet}=10^{42}$\ergps, to $3.7^{+3.3}_{-1.7}$ per cent at $P_{\rm
jet}=10^{44}$\ergps. (Here, the quoted uncertainties 
include all sources of statistical and systematic error discussed 
in the text, including systematic scatter in the 
log\,$M_{\rm BH}-$log\,$\sigma$ relation.)

The existence of such a tight correlation suggests that the Bondi
formalism provides a reasonable description of the accretion process
in these systems, 
despite the fact that the accreting gas contains magnetic fields (\eg
Taylor \etal 2006) and, presumably, has some angular
momentum. Moreover, the similarity of the $P_{\rm jet}$ and $P_{\rm
Bondi}$ values suggests that a significant fraction of the matter
passing through the accretion radius flows all the way down to regions
close to the black hole, where it presumably provides the power source
for the jets. This provides an important constraint on both accretion
and jet formation models. In particular, our results limit the amount
of material that may be lost from the accretion flows en route to the
region of jet formation and requires
that jet formation must be efficient; a few per cent of the energy
associated with the rest mass of material entering the accretion
radius emerges in the jets.

The origin of the jets may be related to the spin of the
central black hole (\eg Rees 1978; Hughes \& Blandford 2003). In that
case, the observed tight correlation between jet power and accretion
rate may suggest a narrow range of black hole spins for the objects in
our sample. The X-ray emitting gaseous halos in the giant elliptical
galaxies studied here are likely to have existed for billions of
years. The growth of the black holes may therefore have been dominated
by gas accretion, which tends to lead to large spin parameters (\eg
Volonteri \etal 2005) and it is possible that the black holes
are rotating at close to their maximal rates. However, the detailed
interaction between accretion, spin-up and jet power remains to be 
explored.

The tight correlation between $P_{\rm Bondi}$ and $P_{\rm jet}$ also
suggests that the accretion flows in the galaxies are stable over the
periods required for gas to flow from the accretion radii to the base
of the jets {\it plus} the times required for the bubbles to be
inflated.  The flow times $t_{\rm flow} \sim r_{\rm A}/c_s$ are of the
order a few $10^4$ to a few $10^{5}$ yr, whereas the bubble ages range
from $10^6-2\times10^7$\yr~(Table~\ref{table:buoy}). Thus, our results
suggest that the accretion flows in the present sample of objects are
approximately stable over timescales of $10^6-10^7$\yr.

The black holes studied here are accreting at rates of a few $10^{-4}$
to a few $10^{-2}$\Msunpyr~and converting a few per cent of their
accreted rest mass into jet energy. The nuclear luminosities
(Pellegrini 2005) are typically 2-4 orders of magnitude lower than the
$P_{\rm jet}$ values, so the systems are `quiescent' in terms of their
nuclear luminosities. (Orientation and obscuration effects will play
some role although are unlikely to modify this conclusion \eg Di
Matteo \etal 2001, 2003; Allen, Di Matteo \& Fabian 2000.) It is
interesting to note that the measured jet powers are comparable to the
bolometric luminosities predicted by standard, radiatively efficient
accretion disk models with $\eta\sim0.1$ (Shakura \& Sunyaev
1973). Thus, it is possible that the ratio of accretion rate/total
emitted power (bolometric luminosity plus jet power) may remain
relatively stable as the accretion mode varies from a quasar to
optically quiescent phase.

Our results have significant implications for models of galaxy
formation and support the idea that feedback from central black holes
is important in shaping the bright end of the galaxy luminosity
function (see \eg Croton \etal 2005 and Bower \etal 2005 
for recent discussions). For our sample, the $observed$ efficiency with 
which energy associated
with the rest mass of accreted matter is fed back into the surrounding
X-ray emitting gas via the jets is similar to that $assumed$ in the
semi-analytic models of \eg Croton \etal (2005). Those models have
been shown to be able to explain the exponential cut off at the bright
end of the galaxy luminosity function and the fact that the most
massive galaxies tend to be bulge dominated systems in clusters,
containing systematically older stars. 

Our results show that the supermassive black holes in elliptical
galaxies are prodigious energy sources, with the bulk of the energy
emerging in the form of relativistic jets rather than radiation.  In
order for this energy to be coupled back into the surrounding matter,
the jets must have a working surface which is provided in these cases
by the X-ray emitting gas. For the elliptical galaxies studied here,
the power fed back into the X-ray emitting gas is, in principle,
sufficient to offset cooling and heat the gas, thereby preventing
further star formation. (Note that Best \etal 2006 apply simple 
scaling arguments to a large elliptical galaxy 
sample to show that the time-averaged power output from radio sources 
can, in principle, balance radiative cooling.)
Since we expect all large galaxies to contain a supermassive black hole,
even at high redshifts, this suggests that a key stage in the
formation of such galaxies will be when they first become large enough
to maintain an X-ray gaseous halo. (This then provides a working
surface for the jets, allowing feedback and preventing further galaxy
growth. Although the jets could have been providing power before this
point, they are unlikely to have had an effective working surface.)
Note, however, that although star formation is truncated when jet
feedback becomes efficient, the black hole may continue to grow via
accretion from the X-ray gas at approximately the Bondi rate.

A closely related topic is that of cooling flows in galaxies and
clusters, and the mechanism responsible for preventing gas cooling to
low temperatures ($kT \approxlt 1$keV).  As stated above, the $P_{\rm
jet}$ values in the galaxies and groups studied here are sufficient to
offset further cooling of the X-ray gas.  It is interesting to
consider extending the correlation between $P_{\rm Bondi}$ and $P_{\rm
jet}$ to larger systems where the cooling of the gas is more
rapid. For the largest cooling flow clusters, the cooling luminosity
exceeds $10^{45}$\ergps. If the black hole masses and accretion rates
in these galaxies are not significantly larger than in NGC507 and M87,
then the jet power is unlikely to be sufficient to balance cooling and
some cooling to very low temperatures and associated star formation
can be expected to occur (see also Fabian \etal 2002). This is
consistent with observations indicating the presence of vigorous star
formation and large molecular gas concentrations in the largest
cooling flow clusters, but relatively little in less X-ray luminous systems
(\eg McNamara \& O'Connell 1992; Allen 1995; Crawford \etal 1999; Edge 2001).

Finally, we note that although systematic uncertainties have been
included in the extrapolation of the gas properties to the Bondi radii
and in our estimation of black hole masses and 
bubble volumes, other sources of
systematic uncertainty remain. For example, the assumption that
$\lambda=0.25$ in the calculation of $P_{\rm Bondi}$ may not be
precise, and our assumption that the bubble ages can be estimated as
$t_{\rm age}$ is probably uncertain at the factor $\sim 2$
level. However, we do not expect such uncertainties to affect the main
conclusions drawn here.

\section*{Acknowledgements}
We thank Roger Blandford, Glenn Morris and Bob Wagoner for helpful
discussions. We also thank the anonymous referee for helpful
comments. This work was supported in part by the U.S. Department of
Energy under contract number DE-AC02-76SF00515. RD and ACF thank PPARC
and the Royal Society for support, respectively. GBT acknowledges
support from the National Aeronautics and Space Administration through
Chandra Award Number GO4-5135X issued by the Chandra X-ray Observatory
Center, which is operated by the Smithsonian Astrophysical Observatory
for and on behalf of the National Aeronautics and Space Administration
under contract NAS8-03060. The National Radio Astronomy Observatory is
a facility of the National Science Foundation operated under a
cooperative agreement by Associated Universities, Inc.

\end{document}